\documentclass[a4paper]{article}

\usepackage{amsmath,graphicx}
 \usepackage{array,multirow,graphicx}
 \usepackage{float}
\usepackage{INTERSPEECH2021}

\usepackage{xcolor}

\title{Significance of Speaker Embeddings and Temporal Context \\ for Depression Detection}
\name{Sri Harsha Dumpala$^{1,2}$, Sebastian Rodriguez$^{1,2}$, Sheri Rempel $^{1,3}$, Rudolf Uher$^{1,3}$, Sageev Oore$^{1,2}$}
\address{
  $^1$Dalhousie University, Canada, $^2$Vector Institute, Canada\\
  $^3$Nova Scotia Health, Halifax, Canada}
\email{\{sriharsha.d, sebastian, sheri.rempel, uher@dal.ca,  sageev@vectorinstitute.ai\}}

\usepackage{xcolor}

\begin{document}

\maketitle
\begin{abstract}
Depression detection from speech has attracted a lot of attention in recent years. However, the significance of speaker-specific information in depression detection has not yet been explored. In this work, we analyze the significance of speaker embeddings for the task of depression detection from speech. Experimental results show that the speaker embeddings provide important cues to achieve state-of-the-art performance in depression detection. We also show that combining conventional OpenSMILE and COVAREP features, which carry complementary information, with speaker embeddings further improves the depression detection performance. The significance of temporal context in the training of deep learning models for depression detection is also analyzed in this paper.
\end{abstract}
\noindent\textbf{Index Terms}: Depression detection, speaker embeddings, temporal context, LSTM, CNN.

\section{Introduction}
\label{sec: intro}


Speech is a complex signal rich in information which includes message, speaker characteristics, emotive state, etc. The speaker characteristics not only provide the identity of the speaker such as gender, age, etc., but are shown to provide important cues about the traits of the speaker such as personality, physical state, likability and pathology \cite{schuller2015survey, dumpala2017improved, narendra2020glottal}. Moreover, speaker-specific information was also considered for emotion classification, and in detection of Alzheimer’s from speech \cite{pappagari2020x, pappagari2020using}.
In this work, we analyze the significance of speaker-specific information in the detection of depression from speech.

Major depressive disorder, also known as depression, is one of the most common mental health disorders and ranks among the health conditions responsible for most disability worldwide
\cite{walker2018prevalence, rehm2019global}. According to World Health Organization \cite{world2015european}, more than 300 million people (around $5\%$ of the global population) suffer from depression, and this number is projected to further increase in the coming years. Early diagnosis of depressive symptoms is crucial in reducing the effects of this disorder. 

As an attempt to aid in depression diagnosis, the problem
of automatically detecting depression using speech has attracted a lot of attention \cite{low2010influence, cummins2015review, ringeval2019avec, valstar2016avec, tao2020spotting}. Recently, the application of deep learning techniques have significantly boosted the performance of depression detection using speech \cite{tasnim2019detecting, ma2016depaudionet, chlasta2019automated, al2018detecting, huang2020exploiting}. Initially deep neural networks (DNNs) with fully-connected layers were considered for depression detection \cite{tasnim2019detecting}. Later, convolutional neural networks (CNNs) and recurrent neural networks with long short-term memory (LSTM)) units were shown to achieve better performance on depression detection \cite{chlasta2019automated, al2018detecting}. Recently, CNN-LSTM and dilated CNN networks were considered for depression detection from speech to achieve state-of-the-art (SOTA) performance \cite{ma2016depaudionet, huang2020exploiting}. 

Speaker-specific information, which provides important cues about speaker traits \cite{schuller2015survey}, was not considered in any of the previous works on depression detection.
In this work, we consider 
speaker-specific information to train CNNs, with multi-sized kernels \cite{sheikh2018sentiment}, and LSTM models for depression detection. 
The main contributions of this work are: 
\begin{itemize}
\item Analyze the significance of speaker embeddings in depression detection from speech.
\item Analyze the significance of temporal context, i.e., number of contiguous segments to be considered, in training deep learning models for depression detection.
\end{itemize}

\section{Related Work}
\label{sec: related_work}
\textbf{Acoustic Representations for Depression Detection:}
Depression is shown to degrade cognitive planning  and psycho-motor functioning thus affecting the human speech production mechanism \cite{cummins2015review}. These effects manifest as variations in the speech voice quality \cite{williamson2014vocal} and several features were proposed to capture these variations in speech for depression detection. Spectral features such as formants,  mel-frequency  cepstral coefficients (MFCCs), prosodic features such as $F_0$, jitter, shimmer and glottal features were initially considered for depression detection \cite{low2010influence, cummins2011investigation, simantiraki2017glottal}. Spectral, prosodic and other voice quality related features extracted using OpenSMILE \cite{eyben2010opensmile} and COVAREP \cite{degottex2014covarep} toolkits were also used for depression analysis \cite{valstar2016avec, al2018detecting}. Further, features developed based on speech articulation such as vocal tract coordination features were considered for depression detection \cite{williamson2014vocal, huang2020exploiting, seneviratne2020extended}. Recently, sentiment and emotion embeddings, representing non-verbal characteristics of speech, are considered for depression severity estimation \cite{dumpala2021estimating}.

To the best of our knowledge, no other studies that we know of have explored the use of speaker-specific information for depression detection. In this work, we consider using speaker embeddings, a representation of speaker-specific information, for depression detection. 

\begin{figure}[!tbp]
\centering
  \centerline{\includegraphics[width=1.0\linewidth]{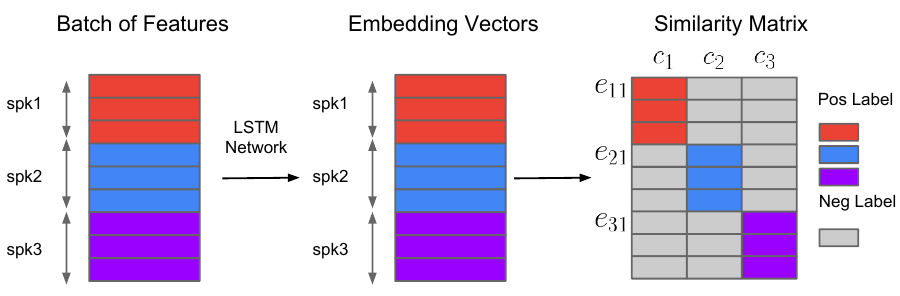}}
  \caption{Overview of the generalized end-to-end loss speaker verification system (Figure taken from \cite{wan2018generalized}). Different colors indicate utterances/embeddings from different speakers.}
  \label{fig:GE2E_overview}
\end{figure}

\noindent\textbf{Speaker Embeddings:}
Speaker embeddings refer to a low-dimensional representation of the speaker-specific characteristics evident in the speech signal \cite{snyder2016deep} \cite{snyder2018x}. Speaker representations were initially based on i-vectors, with a probabilistic linear discriminant analysis (PLDA) back-end  \cite{dehak2010front}. Later, end-to-end deep neural networks based approaches were considered for speaker verification which obtained state-of-the-art performance \cite{snyder2018x, wan2018generalized}. In \cite{snyder2018x}, speaker embeddings, also referred to as x-vectors, were extracted from a time-delay DNN trained for the task of speaker verification. Whereas in \cite{wan2018generalized}, end-to-end LSTM network trained for speaker verification was considered for extracting the speaker embeddings.
In this paper, we consider the generalized end-to-end text-independent speaker verification system (shown in Figure \ref{fig:GE2E_overview}) proposed in \cite{wan2018generalized} to extract the speaker embeddings from speech.


\noindent\textbf{Temporal Context in Depression Detection:}
A few studies have analyzed the effect of the total duration of the audio recording on the depression detection performance \cite{yang2016decision, pampouchidou2016depression, rutowski2019optimizing}. These works have shown that longer the duration, better the performance.
In \cite{yang2016decision, pampouchidou2016depression}, the analysis was performed by considering multiple modalities i.e, audio, visual and text. Whereas in \cite{rutowski2019optimizing}, automatic speech-to-text transcriptions were considered to analyze the effect of duration on depression detection performance. In this work, we directly consider the acoustic features extracted from speech to analyze the effect of varying the number of contiguous speech segments on the performance of LSTM and CNN models trained for depression detection.

\section{Database}
\label{sec: data_details}
Two different depression datasets i.e., DAIC-Woz (corpus of clinical interviews) and FORBOW (spontaneous speech corpus) are considered for analysis in this paper.

\textbf{DAIC-WoZ}: The DAIC-WoZ dataset contains a set of $189$ clinical interviews.
Each interview was conducted between a client and a virtual agent controlled by a human interviewer placed in another location \cite{gratch2014distress}.
The audio recordings of the dataset is divided into train, validation and
test sets, consisting of $107$, $35$, $47$ audio samples, respectively (adopted same partitions as in \cite{valstar2016avec}). Each audio file was labeled with PHQ-8 (Patient Health Questionnaire) score which is in the range of $0-24$ to denote the severity of depression. Audio files with depression score (PHQ-8) $10$ or above are considered as depressed, and those audio files with depression scores below $10$ are considered as non-depressed. Timestamps were provided to each response of the client to the interviewer questions. In this work, each recording is divided into non-overlapping segments of at least $5$ seconds duration. Multiple contiguous responses are combined to form a single segment, if the duration of a response is less than $5$ seconds. A total of $13386$ segments (train: $7255$, valid: $2548$ and test: $3583$ segments) are obtained.

\textbf{FORBOW Dataset}:
Speech data collected as part of the FORBOW (Families Overcoming Risks and Building Opportunities for Well Being) research project \cite{uher2014familial} 
are considered for analysis. Speech samples were collected from $517$ subjects ($390$ mothers and $127$ fathers). In these recordings, parents were asked to talk about their children for five minutes without interruption.
The dataset is divided into train, validation and
test sets, consisting of $352$, $65$, $100$ samples, respectively.
Trained clinical assessors interviewed each participant and scored their current depression severity on the Montgomery and Asberg Depression Rating Scale (MADRS), a validated measure of depression severity \cite{montgomery1979new}.
The range of MADRS scores in this database is $0-21$.
Audio files with depression score (MADRS) $10$ or above are considered as depressed, and those audio files with depressive scores below $10$ are considered as non-depressed.
Each audio recording is divided into non-overlapping segments of $5$ second duration for training and testing the machine learning models. 
A total of $25772$ segments (train: $17524$, valid: $3189$ and test: $5059$ segments) are obtained.
For both datasets, the depression label of a segment is same as the depression label of the overall audio recording.

\begin{figure}[!tbp]
\centering
  \centerline{\includegraphics[width=0.85\linewidth]{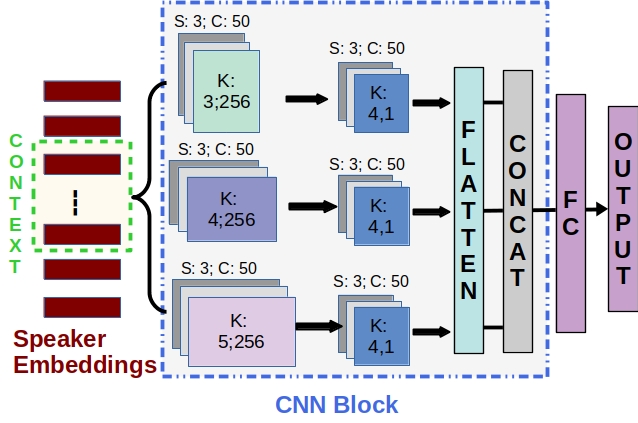}}
  \caption{Network for depression detection using speaker embeddings as input. Same network is considered for OpenSMILE and COVAREP features. FC refers to fully-connected layer.}
  \label{fig:CNN_network}
\end{figure}

\section{Proposed Approach}
\label{sec: Approach}
\subsection{Extraction of Speaker Embeddings}
We consider the generalized end-to-end (GE2E) speaker verification model proposed in \cite{wan2018generalized} to extract the speaker embeddings from speech, for analysis in this work. We provide below a brief review of the GE2E approach (shown in Figure \ref{fig:GE2E_overview}). GE2E training is based on processing a large number of utterances at once, in the form of a batch that contains N speakers, and M utterances from each speaker. 
Each feature vector $x_{ij}$ ($1 \leq j \leq N$ and $1 \leq i \leq M$) represents the features extracted from speaker $j$ utterance $i$.
The features extracted from each utterance $x_{ij}$ is fed into a deep LSTM network with $3$ LSTM layers (with $256$ units each) followed by a fully-connected (FC) layer (with $256$ units). The final fully-connected layer is the embedding layer.
The output of the final layer of the network is denoted as $f(x_{ji};w)$ where $w$ represents parameters of the neural network. The embedding vector (also known as d-vector) is defined as the $L2$ normalization of the final layer output.
\begin{equation*}
e_{ij} = \frac{f(x_{ij}, w)}{\lVert f(x_{ij}, w) \rVert}
\end{equation*}
$C_j$ in Figure \ref{fig:GE2E_overview} refers to the centroid of speaker $j$ obtained by computing the mean of the embedding vectors corresponding to speaker $j$.
In this work, the GE2E network is pre-trained on the task of speaker verification by consolidating $3$ different datasets i.e., LibriSpeech \cite{panayotov2015librispeech}, VoxCeleb1 and VoxCeleb2 \cite{nagrani2017voxceleb} with $1166$ speakers, $1211$ speakers and $5994$ speakers, respectively. Each batch consists of N = $64$ speakers and M = $10$ utterances per speaker. $40$-dimensional MFCCs extracted using a window of size $30$ msec and a step size of $10$ msec are used as the input features. In training, $160$ contiguous frames are randomly selected for each sample. 
This trained GE2E model is then used to extract speaker embeddings at segment-level for the DAIC-WoZ and FORBOW datasets. Each segment is represented using a speaker embedding of dimension $256$. These speaker embeddings are then used to train and test the LSTM and CNN models for depression detection. Note that the GE2E network is not trained on the depression datasets. 


\begin{figure}[!tbp]
\centering
  \centerline{\includegraphics[width=0.77\linewidth]{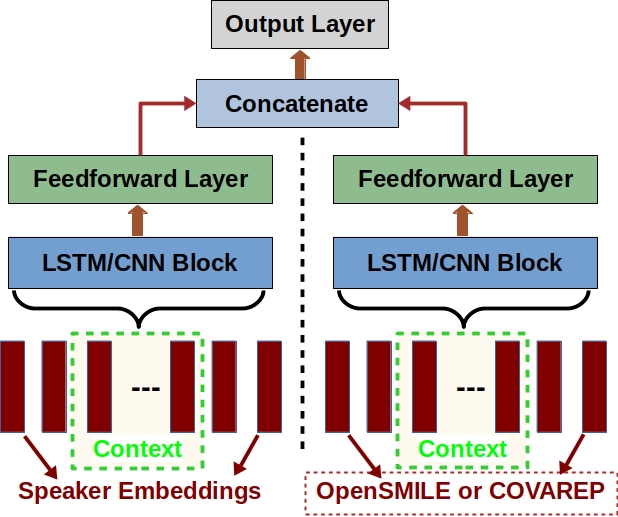}}
  \caption{Network combining speaker embeddings, and OpenSMILE or COVAREP features for depression detection.}
  \label{fig:Comb_network}
\end{figure}

\subsection{Speaker Embeddings for Depression Detection}
We explore the use of both CNN (shown in Figure \ref{fig:CNN_network}) and LSTM networks for depression detection when the speaker embeddings are provided as input.

\noindent\textbf{CNN for Depression Detection (CNN$_D$):} A CNN with multiple kernels, as shown in Figure \ref{fig:CNN_network}, is used for depression detection from the extracted speaker embeddings. The first convolutional layer consists of $3$ different kernels with sizes ($3,D$), ($4,D$) and ($5,D$), respectively. Here, $D$ refers to the length of the input feature vector ($D$ is $256$, $384$ and $444$ for speaker embeddings, OpenSMILE and COVAREP features, respectively). Each kernel consists of $50$ channels. In the second convolutional layer, all kernels are of size $4$ with $50$ channels in each kernel. Outputs from each kernel of the second convolutional layer are flattened and then concatenated before passing through a fully-connected (FC) layer with $100$ units, and then through an output softmax layer with $2$ units.

\noindent\textbf{LSTM for Depression Detection (LSTM$_D$):} Depression detection from the extracted speaker embeddings is also performed by considering an LSTM network. The LSTM network is same as the CNN$_D$ network shown in Figure \ref{fig:CNN_network}, but the CNN block is replaced by an LSTM block, consisting of $2$ LSTM layers with $128$ units each. The output of the LSTM block for the last timestep is passed through the FC layer with $100$ units, and then through an output softmax layer with $2$ units  for obtaining the final decision. 

\noindent\textbf{Baseline FC Network for Depression Detection (DNN$_D$):} A fully-connected deep neural network (DNN) is considered for comparison. This DNN has $3$ hidden layers with $128$, $64$ and $128$ ReLU units, respectively, and a softmax output layer with $2$ units for obtaining the final decision.  

Further, COVAREP \cite{degottex2014covarep, al2018detecting} and OpenSMILE \cite{eyben2010opensmile} features are considered for performance comparison with speaker embeddings. COVAREP and OpenSMILE features are extracted at segment-level to train and test the CNN$_D$ and LSTM$_D$ networks. $384$-dimensional OpenSMILE features representing each segment are obtained by using the $IS09$ configuration from OpenSMILE toolkit. Segment-level COVAREP features ($444$-dimensional) are obtained by computing the higher-order statistics (mean, maximum, minimum, standard deviation, skew, and
kurtosis) of the $74$-dimensional frame-level features (frame-size of $20$ msec and frame-shift of $10$ msec).

\noindent\textbf{Combined Embeddings for Depression (CE$_D$):} We also try combining speaker embeddings with each of OpenSMILE or COVAREP features, respectively (Figure \ref{fig:Comb_network}), for depression detection. As shown in Figure \ref{fig:Comb_network}, the proposed network consists of two branches, one for speaker embeddings and the other for OpenSMILE or COVAREP features. In each branch, the input features are passed through an LSTM (CE$_{DL}$) or CNN (CE$_{DC}$) block and then through a fully-connected (FC) layer ($100$ units). The outputs of the FC layer of each branch are concatenated, and passed through an output layer to get the final decision. Various concatenation techniques (summation, dot product, concatenation and average) are considered. Dot product concatenation gave the best results.

The context in Figures \ref{fig:CNN_network} and \ref{fig:Comb_network} refers to the number of contiguous segments in an audio recording considered to train and test the models. we experiment with temporal contexts of different length to analyze the optimal number of contiguous segments required to train the CNN$_D$ and LSTM$_D$ models for better performance (see Section \ref{ss: Imp_Context}). Note that even though the networks are trained and tested at segment-level, the final accuracy is based on the prediction for the entire audio file. Majority voting is performed on the segment-level decisions to obtain the final decision i.e., depressed or not-depressed. 

\begin{table}
  \caption{{Depression detection performance in terms of $F_1$ and Accuracy (Acc.), when speaker embeddings are considered.}}%
  \label{tab:Spk_emb}
  	\centerline{
    \begin{tabular}{|c|l|c|c|c|c|}
  \hline
  &Model & Context &$F_{1D}$ & $F_{1H}$ & Acc. \\ \hline
  \parbox[t]{2mm}{\multirow{3}{*}{\rotatebox[origin=c]{90}{DAIC}}}  &DNN$_D$ & 1& .32 & .74 & .63 \\ \cline{2-6}
  &CNN$_D$ & 20 & .42 & .77 & .68 \\ \cline{2-6}
  &LSTM$_D$ & 20 & \textbf{.44} & \textbf{.78} & \textbf{.69} \\ \hline
  \parbox[t]{2mm}{\multirow{3}{*}{\rotatebox[origin=c]{90}{FORB.}}} &DNN$_D$ & 1 &.28 & .74 & .65 \\ \cline{2-6}
  &CNN$_D$ &16 & .31 & .79 & .70 \\ \cline{2-6}
  &LSTM$_D$ & 16 & \textbf{.35} & \textbf{.80} & \textbf{.72} \\ \hline
  \end{tabular}}
\end{table}

\begin{table}
  \caption{{Depression detection performance when speaker embeddings (Spk-Emb) are combined with COVAREP (COV) and OpenSMILE (OS) features. CE$_{DD}$, CE$_{DC}$ and CE$_{DL}$ refer to CE$_D$ with DNN, CNN and LSTM blocks, respectively}}
  \label{tab:comb_results}
  	\centerline{
  \begin{tabular}{|c|l|c|c||l|c|c|}
  \hline
  \parbox[t]{1.2mm}{\multirow{10}{*}{\rotatebox[origin=c]{90}{DAIC-WoZ}}}& \multicolumn{3}{|c||}{COVAREP} & \multicolumn{3}{|c|}{(Spk-Emb, COV)} \\ \cline{2-7}
  & & $F_{1D}$/$F_{1H}$ & Acc. & & $F_{1D}$/$F_{1H}$ & Acc. \\ \cline{2-7}
  &DNN$_D$ & .31/.64 & .56 & CE$_{DD}$& .32/.74 & .63 \\ \cline{2-7}
  &CNN$_D$ & .36/\textbf{.71} & \textbf{.61} & CE$_{DC}$& .43/.78 & .69 \\ \cline{2-7}
  &LSTM$_D$ & \textbf{.37}/.69 & .60 &CE$_{DL}$& \textbf{.46/.78} & \textbf{.70} \\ \cline{2-7}
  & \multicolumn{3}{|c||}{OpenSMILE} & \multicolumn{3}{|c|}{(Spk-Emb, OS)} \\ \cline{2-7}
  & & $F_{1D}$/$F_{1H}$ & Acc. & & $F_{1D}$/$F_{1H}$ & Acc. \\ \cline{2-7}
  &DNN$_D$ & .31/.70 & .59 &CE$_{DD}$& .34/.76 & .65 \\ \cline{2-7}
  &CNN$_D$ & .35/.73 & .63 &CE$_{DC}$& .48/.80 & .72 \\ \cline{2-7}
  &LSTM$_D$ & \textbf{.36/.74} & \textbf{.64} &CE$_{DL}$& \textbf{.50/.82} & \textbf{.74} \\ \hline \hline
  \parbox[t]{1.2mm}{\multirow{10}{*}{\rotatebox[origin=c]{90}{FORBOW}}}& \multicolumn{3}{|c||}{COVAREP} & \multicolumn{3}{|c|}{(Spk-Emb, COV)} \\ \cline{2-7}
  & & $F_{1D}$/$F_{1H}$ & Acc. & & $F_{1D}$/$F_{1H}$ & Acc. \\ \cline{2-7}
  &DNN$_D$ & .29/.67 & .59 &CE$_{DD}$& .30/.75 & .67 \\ \cline{2-7}
  &CNN$_D$ & .31/\textbf{.69} & \textbf{.62} &CE$_{DC}$& \textbf{.33/.80} & \textbf{.71} \\ \cline{2-7}
  &LSTM$_D$ & \textbf{.34}/.68 & \textbf{.62} &CE$_{DL}$& .\textbf{33/.80} & \textbf{.71} \\ \cline{2-7}
  & \multicolumn{3}{|c||}{OpenSMILE} & \multicolumn{3}{|c|}{(Spk-Emb, OS)} \\ \cline{2-7}
  & & $F_{1D}$/$F_{1H}$ & Acc. && $F_{1D}$/$F_{1H}$ & Acc. \\ \cline{2-7}
  &DNN$_D$ & .24/.72 & .62 &CE$_{DD}$& .35/.77 & .69 \\ \cline{2-7}
  &CNN$_D$ & \textbf{.28}/.75 & \textbf{.66} &CE$_{DC}$& .39/.82 & .74 \\ \cline{2-7}
  &LSTM$_D$ & .26/\textbf{.76} & \textbf{.66} &CE$_{DL}$& \textbf{.42/.83} & \textbf{.76} \\ \hline
  \end{tabular}}
\end{table}

\noindent\textbf{Training Details:} All the networks were trained using Adam optimizer ($\beta_1=0.9$, $\beta_2=0.99$) with an initial learning rate of $0.0005$. Dropout rate of $0.3$, $0.4$  and $0.3$ was considered for the CNN block, LSTM block and FC layers, respectively to avoid model over-fitting. ReLU activation was used for all the CNN, LSTM and FC layers. Softmax activation for the output layer. All networks were trained for $50$ epochs with a batch size of $128$. Negative log-likelihood (NLL) loss functions was considered to train models. Class weights were set based on the distribution of samples in the train set to alleviate the class imbalance issue during training.
It is to be noted that the same temporal context (number of contiguous segments in a sample) was maintained in the train, validation and testing phase. 

\begin{table}
  \caption{{Depression detection performances comparing proposed approaches with state-of-the-art approaches.}}%
  \label{tab:Compare_SOTA}
  	\centerline{
  \begin{tabular}{|c|l|c|c|c|}
  \hline
  &Approach & $F_{1D}$ & $F_{1H}$ & Acc. \\ \hline
 \parbox[t]{2mm}{\multirow{9}{*}{\rotatebox[origin=c]{90}{DAIC-WoZ}}}
  &Sequence \cite{al2018detecting} & .37 & .69 & .60 \\ \cline{2-5}
  &eGeMAPS \cite{huang2019investigation} & .31 & .70 & .59 \\ \cline{2-5}
  &FVTC-MFCC \cite{huang2020exploiting} & .38 & .78 & .67 \\ \cline{2-5}
  &FVTC-FMT \cite{huang2020exploiting} & .41 & .78 & .68 \\ \cline{2-5}
  &CNN$_D$(Spk-Emb) & .42 & .77 & .68 \\ \cline{2-5}
  &LSTM$_D$(Spk-Emb) & .44 & .78 & .69 \\ \cline{2-5}
  &CE$_{DC}$(Spk-Emb, OS) & .48 & .80 & .72 \\ \cline{2-5}
  &CE$_{DL}$(Spk-Emb, OS) & \textbf{.50} & \textbf{.82} & \textbf{.74} \\ \hline \hline
   \parbox[t]{2mm}{\multirow{9}{*}{\rotatebox[origin=c]{90}{FORBOW}}}
  &Sequence \cite{al2018detecting} & .34 & .68 & .62 \\ \cline{2-5}
  &eGeMAPS \cite{huang2019investigation} & .25 & .73 & .63 \\ \cline{2-5}
  &FVTC-MFCC \cite{huang2020exploiting} & .28 & .76 & .67 \\ \cline{2-5}
  &FVTC-FMT \cite{huang2020exploiting} & .33 & .77 & .69 \\ \cline{2-5}
  &CNN$_D$(Spk-Emb) & .31 & .79 & .70 \\ \cline{2-5}
  &LSTM$_D$(Spk-Emb) & .35 & .79 & .71 \\ \cline{2-5}
  &CE$_{DC}$(Spk-Emb, OS) & .39 & .81 & .73 \\ \cline{2-5}
  &CE$_{DL}$(Spk-Emb, OS) & \textbf{.42} & \textbf{.83} & \textbf{.76} \\ \hline
\end{tabular}}
\end{table}

\section{Experimental Results}
\label{sec: results}
Depression detection performance scores when speaker embeddings are considered to train DNN$_D$, CNN$_D$ and LSTM$_D$ models are given in Table \ref{tab:Spk_emb}.  It can be observed from Table \ref{tab:Spk_emb} that the LSTM and CNN models achieve better performance when compared to DNN on both DAIC-WoZ (DAIC) and FORBOW (FORB.) datasets. $F_{1D}$ and $F_{1H}$ are $F_1$ scores of depressed and healthy classes, respectively. In this work, Acc. refers to the weighted accuracy of the two classes i.e., depressed and non-depressed (healthy).

Table \ref{tab:comb_results} shows the depression detection performance when speaker embeddings are combined with each of OpenSMILE (Spk-Emb, OS) or COVAREP (Spk-Emb, COV) features, respectively. It can be observed from Tables \ref{tab:Spk_emb} and \ref{tab:comb_results} that the models trained on speaker embeddings outperform the models trained on COVAREP or OpenSMILE features for both DAIC-WoZ and FORBOW datasets. It can also be observed that combining speaker embeddings with OpenSMILE or COVAREP features further improves the depression detection performance. This shows that the speaker embeddings carry complementary information when compared to OpenSMILE or COVAREP features. Moreover, the LSTM$_D$ and CNN$_D$ outperformed the DNN$_D$ in all conditions, with the LSTM$_D$ performing better or similar to the CNN$_D$ models. For COVAREP and OpenSMILE features, temporal context of $20$ and $16$ is considered for DAIC-WoZ and FORBOW datasets, respectively.

Table \ref{tab:Compare_SOTA} compares the performance of the proposed approaches with state-of-the-art (SOTA) approaches. It can be observed from Table \ref{tab:Compare_SOTA} that the models trained on speaker embeddings perform better than the SOTA approaches for depression detection using speech. It can also be observed that the depression detection performances obtained by combining speaker embeddings with the OpenSMILE features (Spk-Emb, OS) outperform the SOTA approaches.

We also analyzed the effectiveness of the extracted speaker embeddings for the task of speaker classification.
 DAIC-WOZ and FORBOW datasets consist of audio recordings corresponding to $189$ and $517$ speakers, respectively. 
For each speaker, $25$ and $15$ non-overlapping segments were randomly selected to form the train and test sets for that speaker, respectively. 
A logistic regression classifier (with no hidden layers) was trained for the task of speaker classification ($189$ and $517$ class classification for DAIC-WoZ and FORBOW datasets, respectively).
On the test sets, equal error rates (EER) of $1.29$ and $1.69$ are obtained for DAIC-WoZ and FORBOW datasets, respectively.
These low EER values show that the extracted speaker embeddings, used in this work, carry speaker-specific information.




\begin{figure}
\centering
  \centerline{\includegraphics[width=0.86\linewidth]{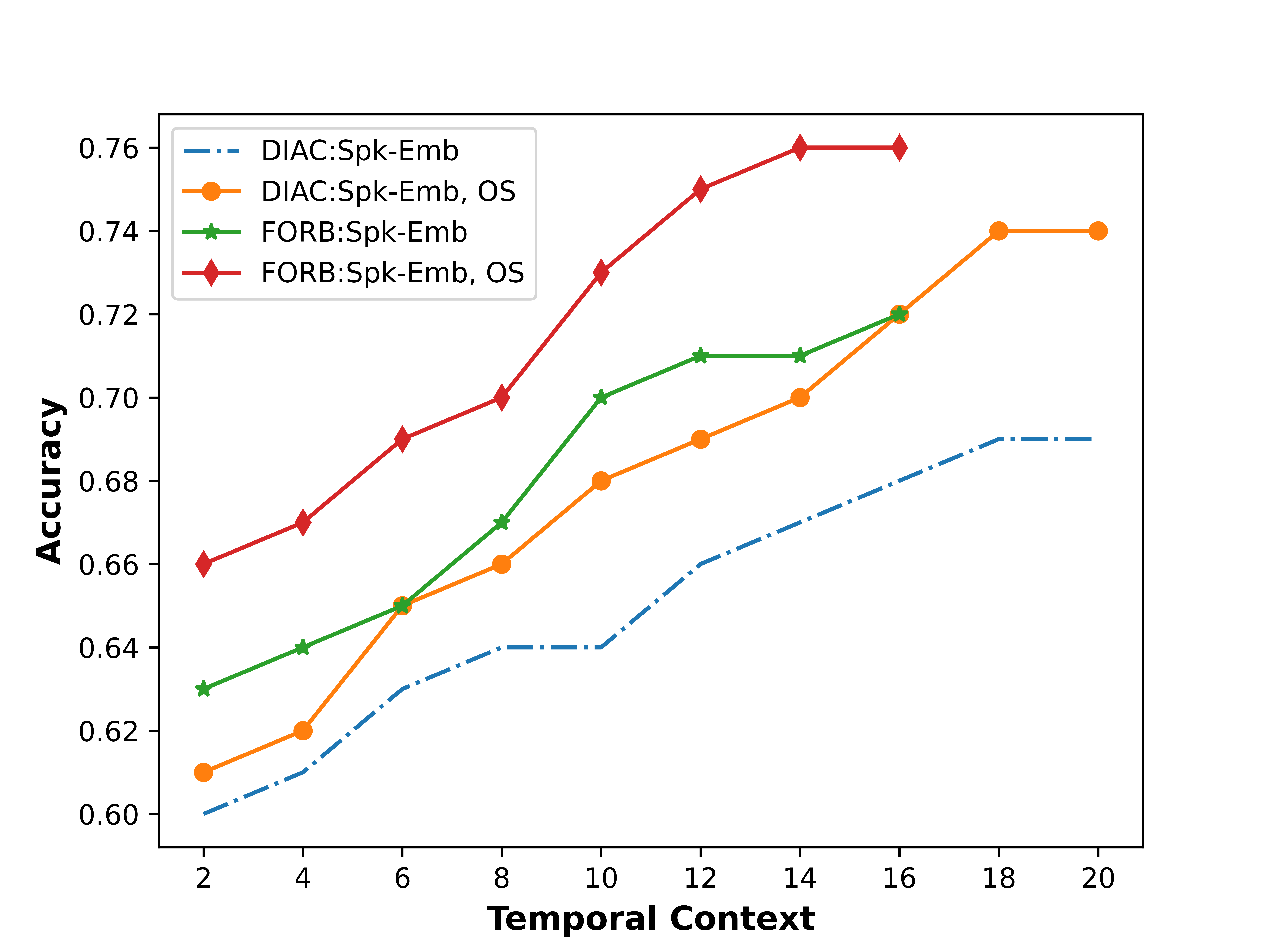}}
  \caption{Performance (Accuracy) of the LSTM$_D$ (Spk-Emb) and CE$_{DL}$ (Spk-Emb, OS) for depression detection when the length of the context is varied from 4 up to 
  the entire length of the shortest example in the test set, i.e. 16 and 20 respectively.}
  \label{fig:context}
\end{figure}

\subsection{Temporal Context in Depression Detection}
\label{ss: Imp_Context}
Figure \ref{fig:context} shows the depression detection performance on the DAIC-Woz (DAIC) and FORBOW (FORB) datasets when different temporal contexts are considered. Two different input configurations are used for each data set: one uses only Speaker Embeddings, and the other combines Speaker Embeddings and OpenSMILE (Spk-Emb, OS) features. Note that it does not make sense to use a context longer than the shortest speech samples in the test set, which are $16$ and $20$ segments in FORBOW and DAIC respectively. Thus our temporal contexts range from roughly $20$ seconds ($4$ contiguous segments) to $80$ or $100$ seconds ($16$ or $20$ contiguous segments).

For all cases, as we increase the temporal context, the depression detection performance tends to improve until saturation. For example, Figure \ref{fig:context} shows that for the CE$_{DL}$ model trained using combined speaker embeddings and OpenSMILE on FORBOW dataset (i.e., FORB:Spk-Emb, OS), as we increase the temporal context up to $16$ segments, the performance of the CE$_{DL}$ improves to an accuracy of $0.76$. 
This indicates that the temporal relationship existing in the features provide important cues for depression detection.


An important area of future work, as larger data sets become available, will be to investigate the relationship of overall duration of the audio samples and the context size at which performance saturates.


\section{Summary}
\label{sec: summary}
In this work, the significance of speaker embeddings for the task of depression detection from speech was analyzed. Experimental results show that the speaker embeddings provide important cues for depression detection. Experimental results also showed that combining speaker embeddings with OpenSMILE features achieves state-of-the-art performance on depression detection. Further, the experimental results show that as we increase the temporal context (i.e., number of contiguous segments considered to train and test deep learning models), the depression detection performance improves.



\bibliographystyle{IEEEtran}
\bibliography{main.bib} 

\end{document}